\documentclass[a4paper]{article}
\usepackage{graphicx}
\usepackage{amsmath}
\begin{document}
\begin{center}
{\bf \Large
Heider Balance in Human Networks
}\\[5mm]

{\large
P. Gawro{\'n}ski and K. Ku{\l}akowski
}\\[3mm]

{\em
Department of Applied Computer Science,
Faculty of Physics and Applied Computer Science,
AGH University of Science and Technology\\
al. Mickiewicza 30, PL-30059 Krak\'ow, Poland
}

\bigskip
{\tt kulakowski@novell.ftj.agh.edu.pl}

\bigskip
\today
\end{center}

\begin{abstract}

Recently, a continuous dynamics was proposed to simulate dynamics of interpersonal relations
in a society represented by a fully connected graph. Final state of such a society was found
to be identical with the so-called Heider balance (HB), where the society is divided into two 
mutually hostile groups. In the continuous model, a polarization of opinions was found in HB. 
Here we demonstrate that the polarization occurs also in Barab\'asi-Albert networks, where the 
Heider balance is not necessarily present. In the second part of this work we demonstrate 
the results of our formalism, when applied to reference examples: the Southern women and the 
Zachary club.
\end{abstract}



\section{Introduction}

 The Heider balance \cite{h46,hei2,hara,dor1,wt} is a final state of personal relations 
between members of a society, reached when these relations evolve according to 
some dynamical rules. The relations are assumed to be symmetric, and they can be friendly 
or hostile. The underlying psycho-sociological mechanism of the rules is an attempt of 
the society members to remove a cognitive dissonance, which we feel when two of our friends 
hate each other or our friend likes our enemy. As a result of the process, the society 
is split into two groups, with friendly relations within the groups and hostile 
relations between the groups. As a special case, the size of one group is zero, 
i.e. all hostile relations are removed. HB is the final state if each member 
interacts with each other; in the frames of the graph theory, where the problem is formulated, 
the case is represented by a fully connected graph.

Recently a continuous dynamics has been introduced to describe the time evolution of the 
relations \cite{my1}. In this approach, the relations between nodes $i$ and $j$were 
represented by matrix elements $r(i,j)$, which were real numbers, friendly ($r(i,j)>0$) or
hostile ($r(i,j)<0)$. As a consequence of the
continuity, we observed a polarization of opinions: the absolute values of the matrix 
elements $r(i,j)$ increase. Here we continue this discussion, but the condition of maximal 
connectivity is relaxed, as it could be unrealistic in large societies. 
The purpose of first part of this work is to demonstrate, that even if HB is not 
present, the above mentioned polarization remains true. In Section II we present 
new numerical results for a society of $N=100$ members, represented by Barab\'asi-Albert (BA)
network \cite{ab}. Although this size of considered social structure is rather small, it is 
sufficient to observe some characteristics which are different than those in the exponential 
networks. In second part (Section III) we compare the results of our equations of motion with 
some examples, established in the literature of the subject. The Section is closed by final 
conclusions.

\section{Calculations for Barab\'asi-Albert networks}

The time evolution of $r(i,j)$ is determined by the equation of motion \cite{my1}

\begin{equation}
\frac{dr(i,j)}{dt}=\Big\{1-\Big(\frac{r(i,j)}{R}\Big)^2\Big\}\sum_k r(i,k) r(k,j)
\end{equation}
where $R$ is a sociologically justified limitation on the absolute value of $r(i,j)$ \cite{my1}. 
Here $R=5.0$. Initial values of $r(i,j)$ are random numbers, uniformly distributed in the range 
$(-0.5,0.5)$.
The equation is solved numerically with the Runge-Kutta IV method with variable length of timestep
\cite{RK4}, simultaneously 
for all pairs $(i,j)$ of linked nodes. The method of construction of BA networks was described
in \cite{MK}. The connectivity parameter is selected to be $M=7$, because in this case the 
probability $p(M)$ of HB has a clear minimum for BA networks of $N=100$ nodes, and 
$p(M=7)\approx 0.5$ (see Fig. 1). This choice of $M$ is motivated by our aim to falsify the result on the 
polarization of opinions. This polarization was demonstrated \cite{my1} to be a consequence of HB;
therefore, the question here is if it appears also when HB is not present. An example of 
time evolution of such a network is shown in Fig. 2.

Our result is that the polarization is present in all investigated cases. As time 
increases,
the distribution of $r(i,j)$ gets wider and finally it reaches a stable shape, with two large peaks 
at $r(i,j)\approx\pm R$ and one smaller peak at the centre, where $r(i,j)\approx 0$.
In Fig. 3, we show a series of histograms of $r(,j)$ in subsequent times (A-E).
Particular networks differ quantitatively with respect to the heights of the peaks, but these 
differences are small. 

We note here that when some links are absent, the definition of HB should be somewhat relaxed,
because some other links, which do not enter to any triad $(i,j,k)$, will not evolve at all. 
Therefore we should admit that some negative relations survive within a given group. We classify a
final state of the graph as HB if there are no chains of friendly relations between the subgroups. 
On the other hand, more than two mutually hostile subgroups can appear. These facts were recognized
already in literature \cite{hara,wt}. Surprisingly enough, subgroups of $1<N<97$ nodes are never 
found in our BA networks. On the contrary, in the exponential networks groups of all sizes
were detected. In Figs. 4 and 5 we show diagrams for BA networks and exponential networks, respectively.
Each point at these diagrams marks the value of $r(i,j)$ and the size of the subgroup which
contains nodes $(i,j)$. Links between different subgroups are omitted. We see that for BA
networks (Fig.4), the lowest value of $N$ is 97. The remaining three nodes are linked with all
other nodes by hostile relations. 

\section{Examples}

In Ref. \cite{my1}, an example of polarization of opinions on the lustration law in Poland in 1999
was brought up. The presented statistical data \cite{cbos} displayed two maxima at negative and
positive opinions and a lower value at the centre of the plot. In our simulations performed for
fully connected graphs \cite{my1}, the obtained value for the center was zero. However, it is clear 
that in any group larger than, say, 50 persons some interpersonal relations will be absent. Taking this
into account, we can claim than the statistical data of \cite{cbos} should be compared to the
results discussed here rather than to those for a fully connected graph. Here we reproduce a 
peak of the histogram at its centre, on the contrary to the results in \cite{my1}. This fact
allows to speak on a qualitative accordance of the results of our calculations with the statistical 
data of \cite{cbos}.

Next example is the set of data of the attendance of 18 'Southern women' in local meetings in 
Natchez, Missouri, USA in 1935 \cite{free}. These data were used to compare 21 methods of finding 
social groups. The results were analysed with respect to their consensus, and ranked with consensus
index from 0.543 (much worse than all others) to 0.968. To apply our dynamics, we use the 
correlation function $<p(i,j)>-<p(i)><p(j)>$ as initial values of $r(i,j)$. Our method produced 
the division (1-9) against (10-18), what gives the index value 0.968. As a by-product, the method 
can provide the time dynamics of the relations till HB and, once HB is reached, the leadership within 
the cliques \cite{bl}. We should add that actually, we have no data on the possible friendship or 
hostility between these women, then the interpretation of these results should be done with care.

Last example is the set of data about a real conflict in the Zachary karate club \cite{za,bonet,gir}.
The input data are taken from \cite{wbpg}. All initial values of the matrix elements are reduced 
by a constant $\epsilon$ to evade the case of overwhelming friendship. The obtained splitting of 
the group is exactly as observed by Zachary: (1-8,11-14,17,18,20,22) against 
(9,10,15,16,19,21,23-34). These results were checked not to vary for $\epsilon$ between 1.0 
and 3.0. The status of all group members can be obtained with the same method as in the previous 
example.

To conclude, the essence of Eq. (1) is the nonlinear coupling between links $r(i,j)$, which produces the positive 
feedback between 
the actual values of the relations and their time evolution. We should add that the idea of such 
a feedback is not entirely new.
It is present, for example, in Boltzmann-like nonlinear master equations applied to behavioral
models \cite{hlb}. On the contrary, it is absent in later works on formal theory of social influence 
\cite{cons}. On the other hand, the theories of status \cite{bl}
are close to the method of transition matrix, known in non-equilibrium statistical mechanics 
\cite{re}.

\newpage

\begin{figure}
\includegraphics[angle=-90,width=.8\textwidth]{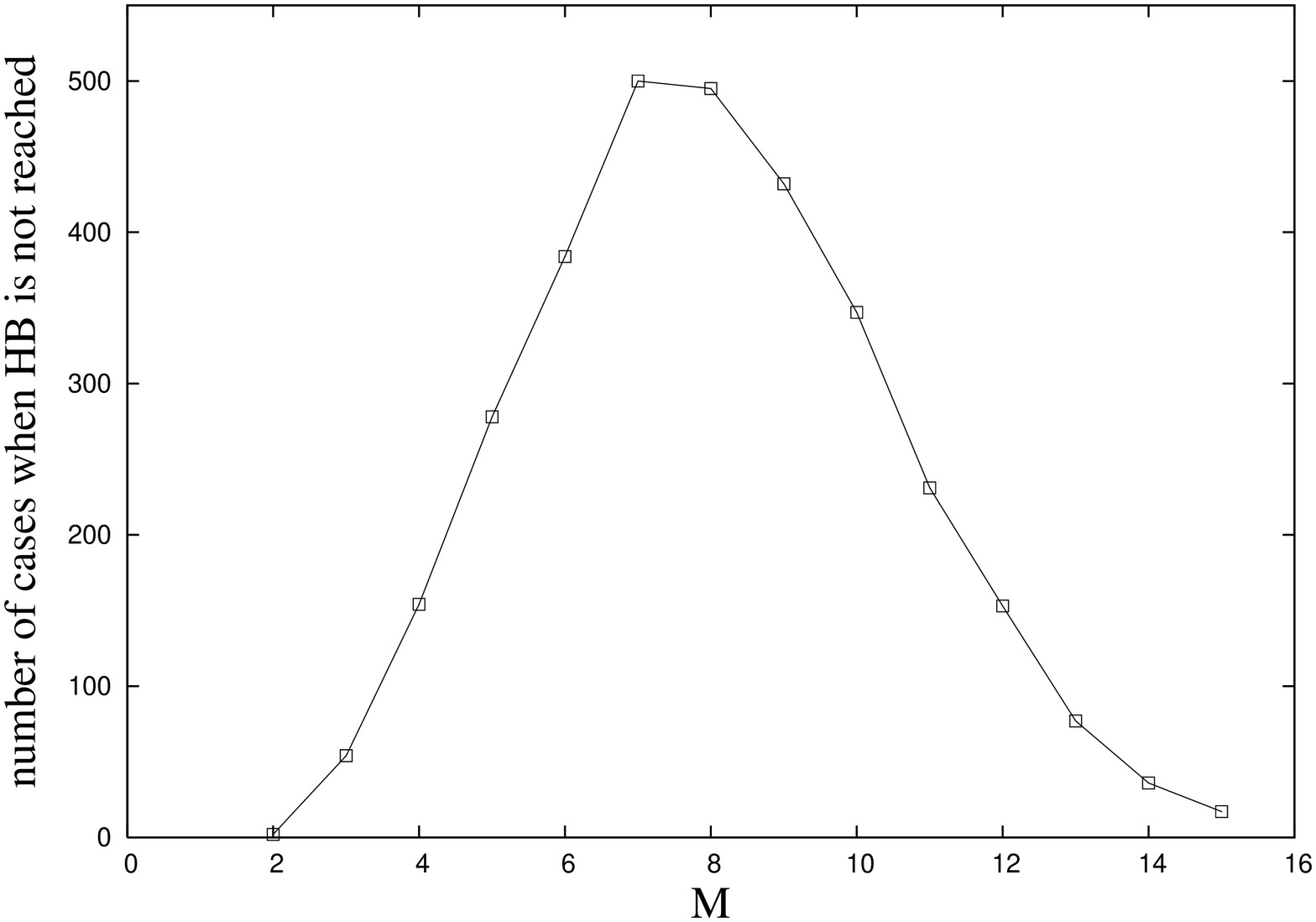}
\caption{The number of networks per 1000 trials, where HB is not reached.}
\label{fig1}
\end{figure}

\begin{figure}
\includegraphics[angle=-90,width=.8\textwidth]{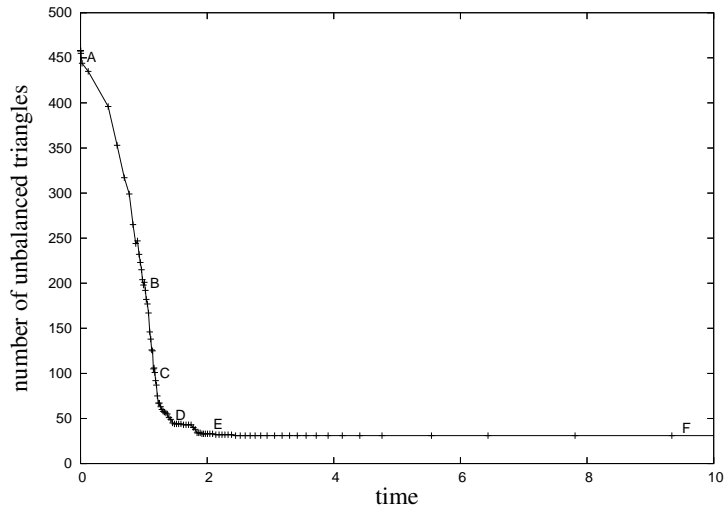}
\caption{The number of unbalanced triangles of nearest neighbours $(i,j,k)$ against time.
Six successive times for Fig. 3 are marked with labels A-F.}
\label{fig2}
\end{figure}

\begin{figure}
\includegraphics[angle=270,width=7.5cm]{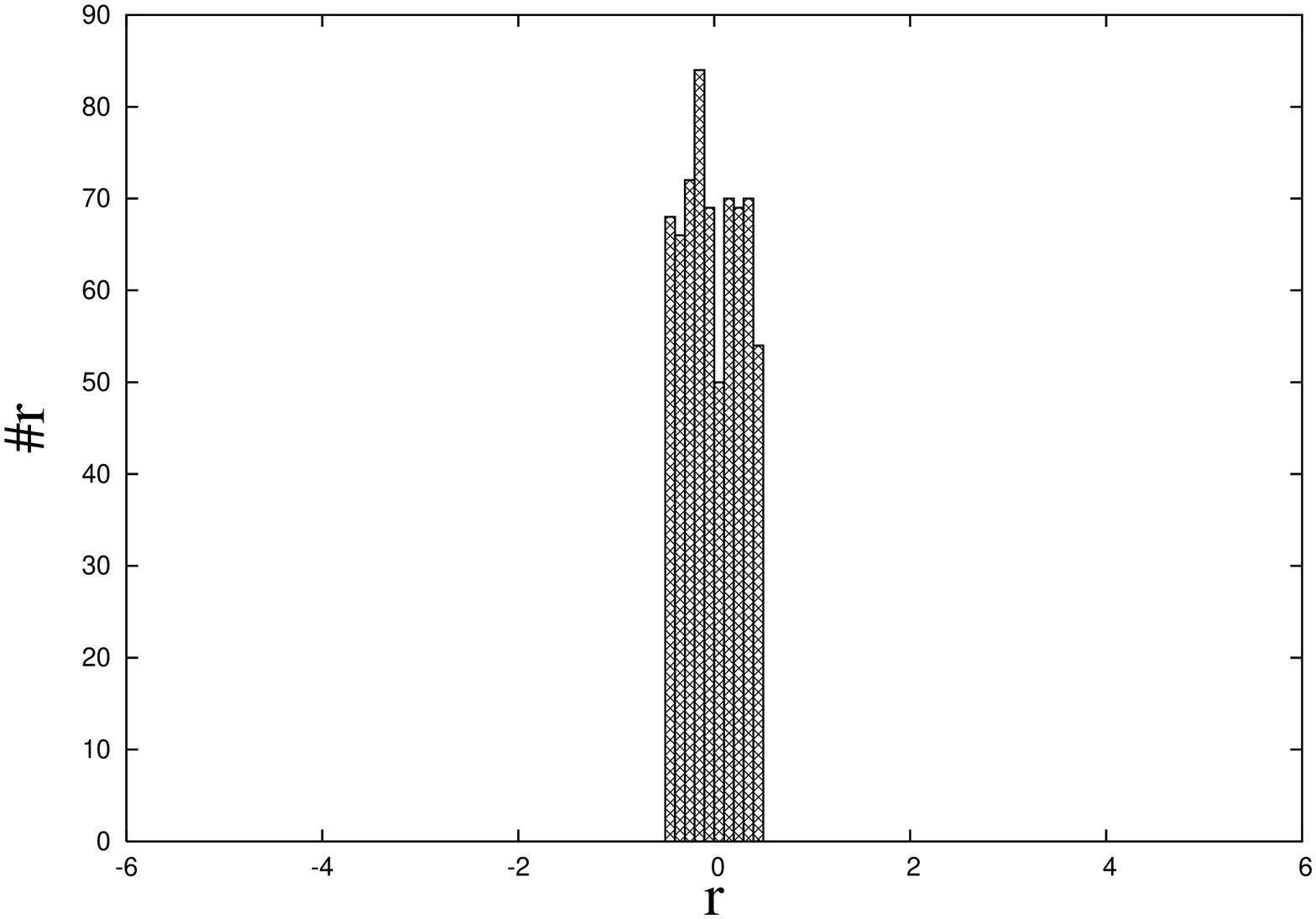}
\includegraphics[angle=270,width=7.5cm]{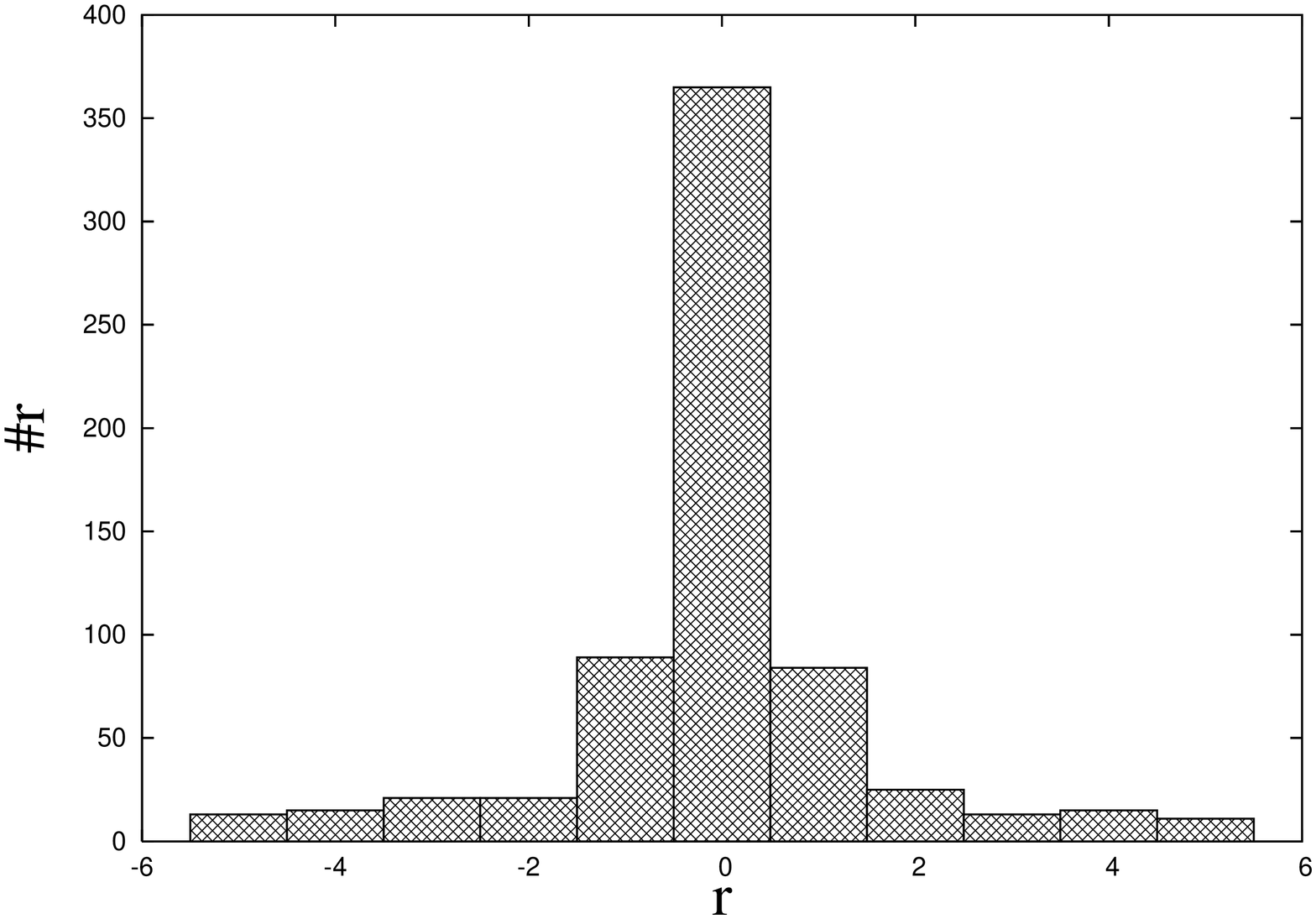}
\includegraphics[angle=270,width=7.5cm]{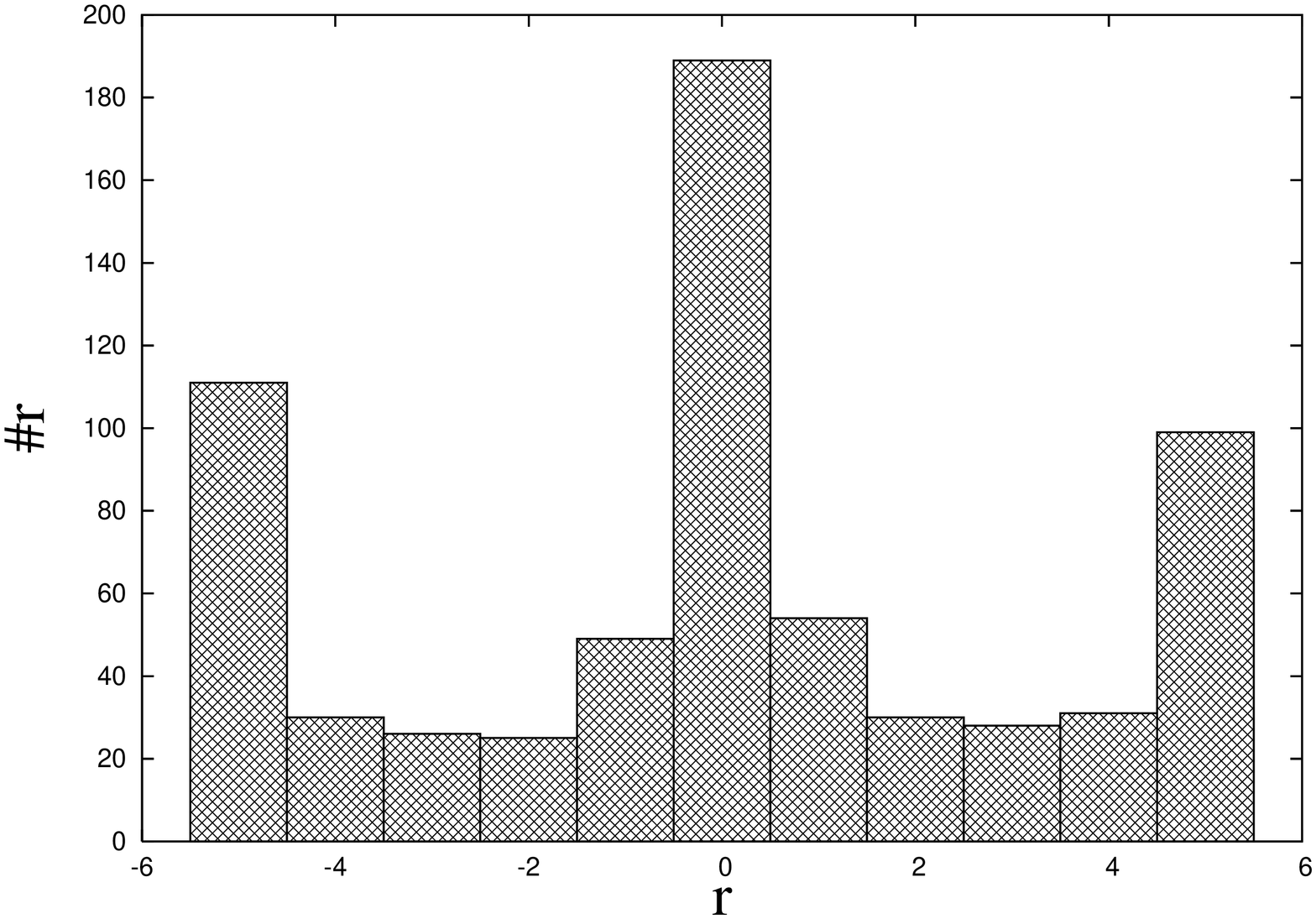}
\includegraphics[angle=270,width=7.5cm]{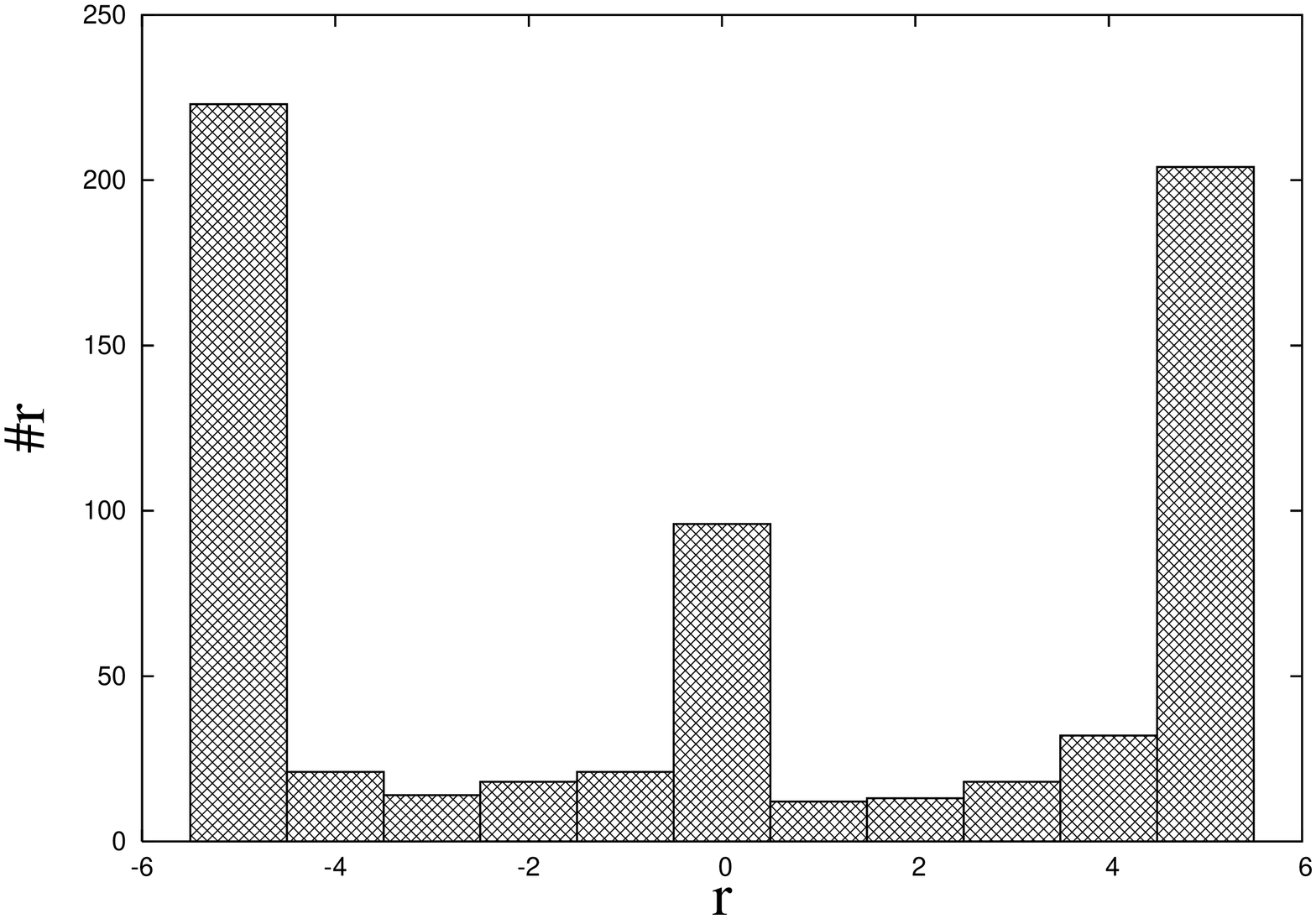}
\includegraphics[angle=270,width=7.5cm]{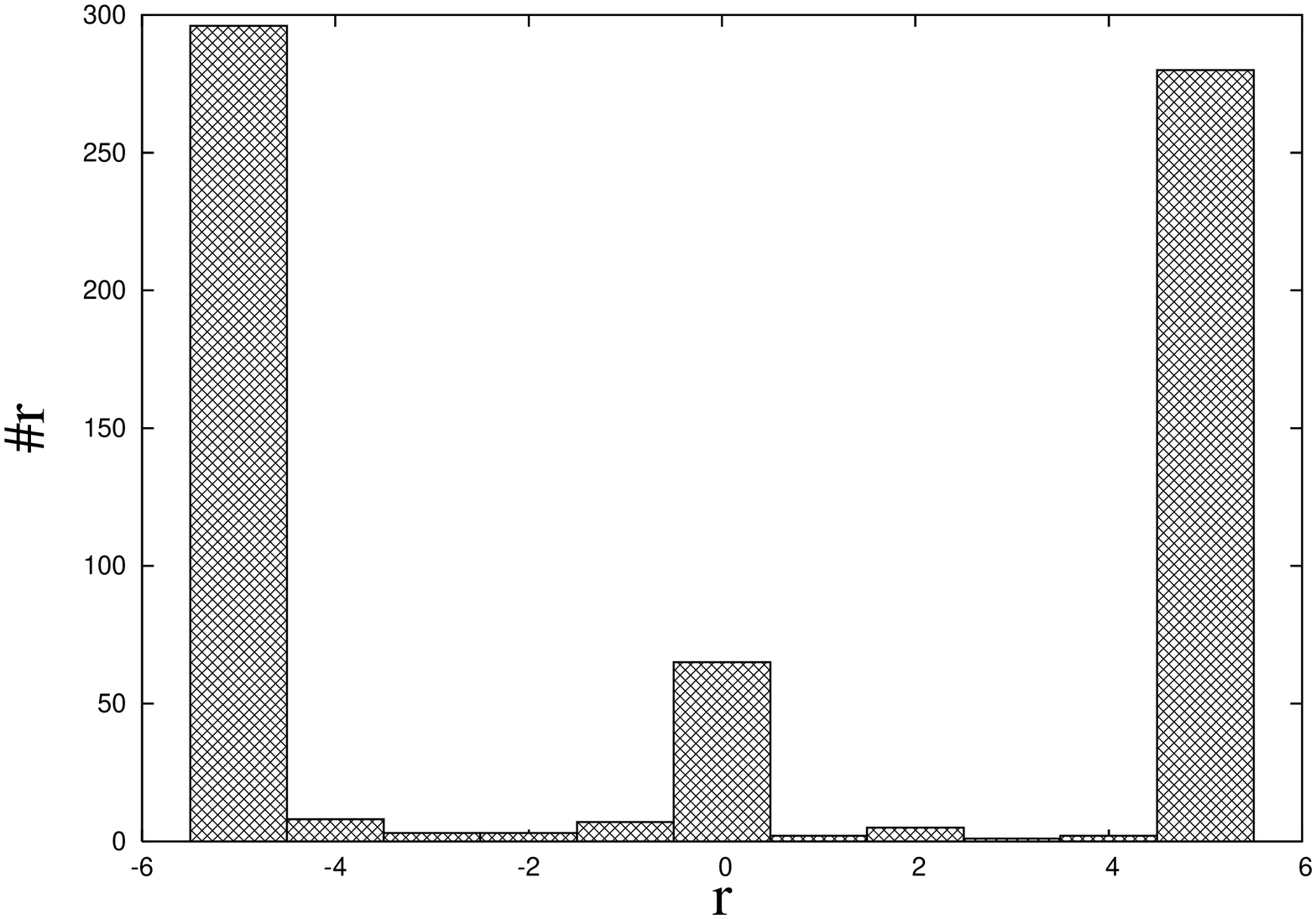}
\includegraphics[angle=270,width=7.5cm]{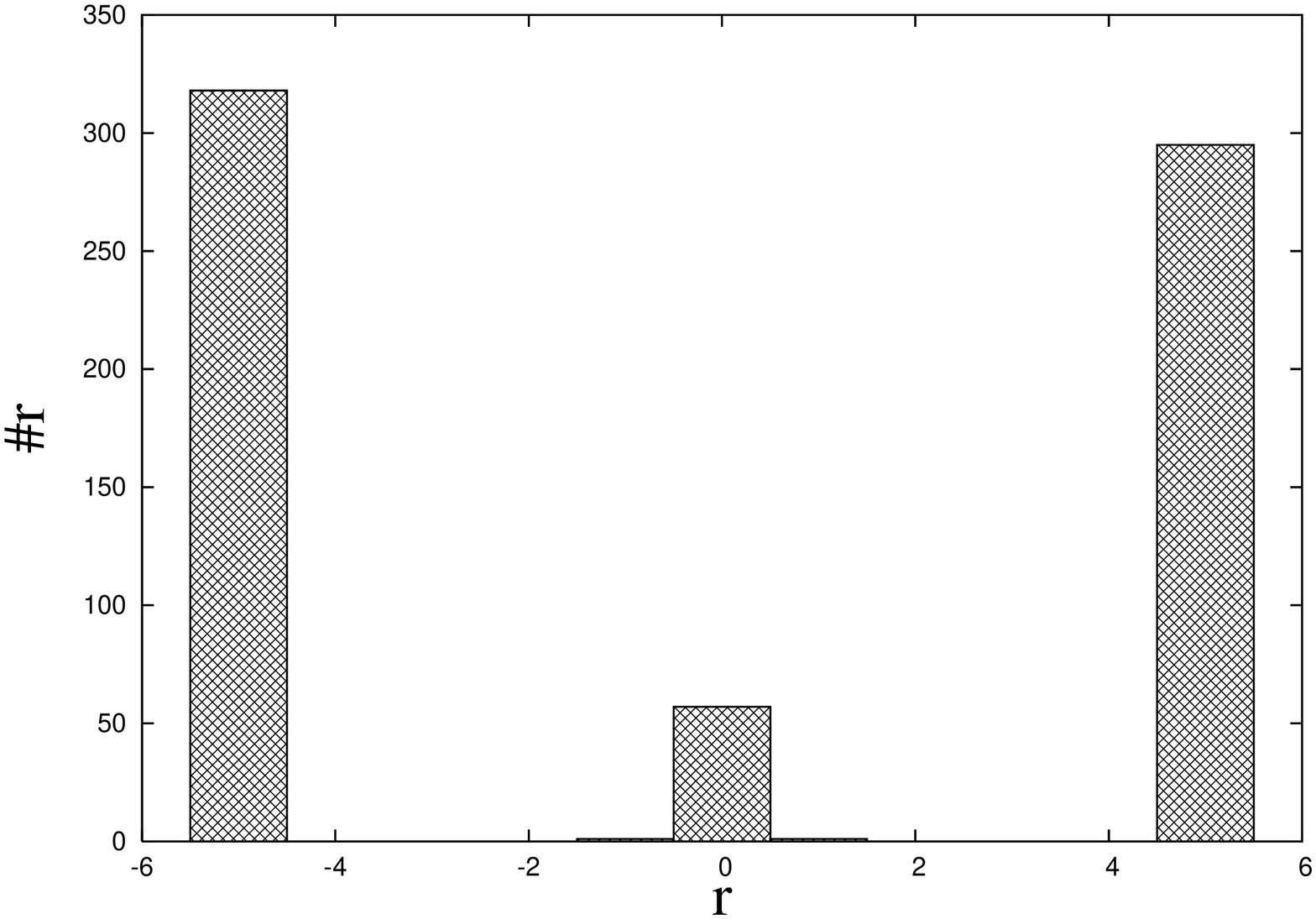}
\caption{Time evolution of the histogram of the matrix elements $r(i,j)$.}
\label{fig3a}
\end{figure}

\begin{figure}
\includegraphics[angle=-90,width=.8\textwidth]{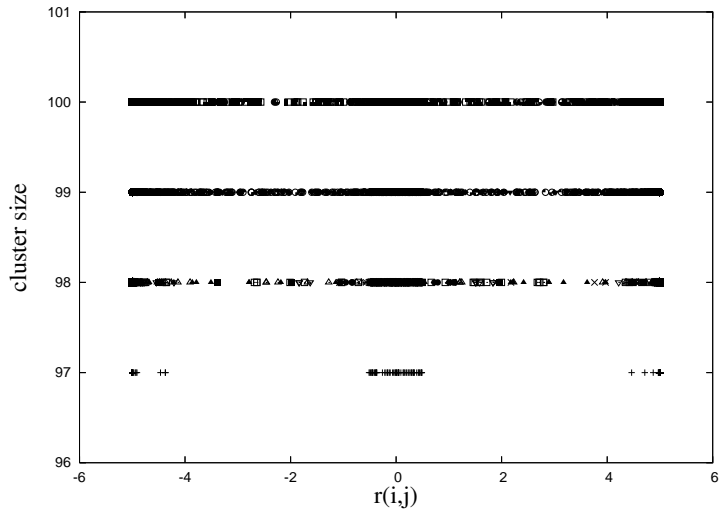}
\caption{Diagram of values of $r(i,j)$ and the group size for BA networks.}
\label{fig4}
\end{figure}

\begin{figure}
\includegraphics[angle=-90,width=.8\textwidth]{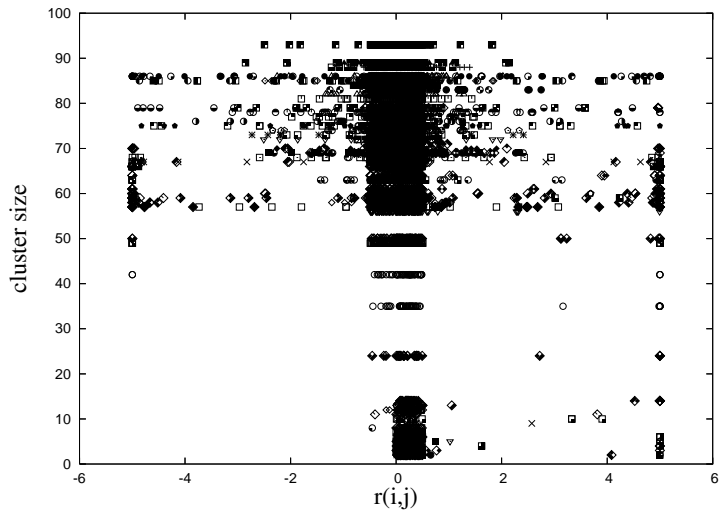}
\caption{Diagram of values of $r(i,j)$ and the group size for exponential networks.}
\label{fig5}
\end{figure}

\end{document}